\documentclass[twocolumn,preprintnumbers,nofootinbib,amsmath,amssymb]{revtex4}

\usepackage{amsmath}
\usepackage{graphicx}
\usepackage{dcolumn}
\usepackage{bm}

\begin{document}


\title{Reconstruction of the Primordial Power Spectrum using
  Temperature and Polarisation Data from Multiple Experiments}

\author{Gavin Nicholson}
 \email{gavin.nicholson05@imperial.ac.uk}
\author{Carlo R. Contaldi}%
 \email{c.contaldi@imperial.ac.uk}
\affiliation{%
Theoretical Physics, Blackett Laboratory, Imperial College, Prince
 Consort Road, London, SW7 2BZ, U.K.}%

\date{\today}

\begin{abstract}
  We develop a method to reconstruct the primordial power spectrum,
  $P(k)$, using both temperature and polarisation data from the joint
  analysis of a number of Cosmic Microwave Background (CMB)
  observations. The method is an extension of the Richardson-Lucy
  algorithm, first applied in this context by Shafieloo \& Souradeep
  \cite{Shafieloo:2003gf}. We show how the inclusion of polarisation
  measurements can decrease the uncertainty in the reconstructed power
  spectrum. In particular, the polarisation data can constrain
  oscillations in the spectrum more effectively than total intensity
  only measurements.  We apply the estimator to a compilation of
  current CMB results. The reconstructed spectrum is consistent with
  the best--fit power spectrum although we find evidence for a `dip'
  in the power on scales $k\approx 0.002$ Mpc$^{-1}$. This feature
  appears to be associated with the WMAP power in the region $18\le
  \ell \le 26$ which is consistently below best--fit models. We also
  forecast the reconstruction for a simulated, Planck--like
  \cite{planck} survey including sample variance limited polarisation
  data.
\end{abstract}

\maketitle

\section{Introduction}

With increasingly precise measurements being made of the CMB (Cosmic
Microwave Background) or LSS (Large Scale Structure) it becomes
progressively more important to determine how new observations
could yield the greatest insight into processes occurring in the early
universe. One such observable is the primordial power spectrum of
curvature perturbations $\Phi({\vec k})$,
\begin{equation}
 P(k) \equiv \frac{k^3}{2\pi^2}\delta^3(\vec{k}-\vec{k'})\langle\Phi(\vec{k})\Phi^*(\vec{k'})\rangle , 
\end{equation}
where $k$ is the wavenumber. 

A generic prediction of the simplest inflationary models is that the
density perturbations should be adiabatic and near scale invariant. In
these models the spectrum takes the form of a power law $\propto
k^{n_s-1}$. Current limits on this parametrisation place the spectral
index $n_s \approx 0.963$ \cite{wmap5}. More complex inflationary
models such as those with features on the potential
\cite{Adams:2001vc,Wang:2002hf,Hunt:2004vt,Joy:2007na,Hunt:2007dn,Pahud:2008ae,Lerner:2008ad},
a small number of $e$-folds
\cite{Contaldi:2003zv,Powell:2006yg,Nicholson:2007by}, or other exotic
inflationary models
\cite{Lesgourgues:1999uc,Feng:2003zua,Mathews:2004vu,Jain:2008dw,Romano:2008rr}
can modify $P(k)$ in a manner not compatible with the simple power law
description. There have been many parametric searches for the features
produced by these models
\cite{Bridle:2003sa,Contaldi:2003zv,Parkinson:2004yx,Sinha:2005mn,Sealfon:2005em,Mukherjee:2005dc,Bridges:2005br,Bridges:2006zm,Covi:2006ci,Joy:2008qd,Verde:2008zz},
although none have proved conclusive. On the other hand, there have
been tantalising hints of anomalous features in the data, for example
after the first year WMAP results were released, there were strong
indications of a cut-off in $P(k)$ on large scales. With subsequent
data releases the significance of this feature has been
reduced, although future observations of the polarisation of the CMB
may provide more conclusive evidence \cite{Nicholson:2007by}.

A more thorough search for features in $P(k)$ would be one in which
the theoretical model is independent from the reconstruction. Methods
such as the Richardson-Lucy deconvolution
\cite{Shafieloo:2003gf,Shafieloo:2006hs,Shafieloo:2007tk}, cosmic
inversion
\cite{Matsumiya:2001xj,Kogo:2003yb,Kogo:2004vt,Nagata:2008tk,Nagata:2008zj},
and other non-parametric approaches
\cite{Hannestad:2000pm,Wang:2000js,Bridle:2003sa,Mukherjee:2003cz,Mukherjee:2003ag,Hannestad:2003zs,TocchiniValentini:2004ht,Leach:2005av,Paykari:2009ac}
all attempt to overcome theoretical bias. With current computing power
these techniques are generally limited to recovering the spectrum for
one set of assumed cosmological parameters. This allows one to use a
fiducial CMB photon transfer function to integrate the primordial
curvature perturbation into today's photon distribution
perturbation. However this assumption hides a significant degeneracy
between features in the primordial power spectrum and the physical
parameters which determine the height and position of acoustic peaks
in the CMB when using only total intensity data in the reconstruction
process \cite{Hu:2003vp}. As such it is not clear what the
significance of any features found in the reconstructed $P(k)$ should
be. Adding polarisation information into the reconstruction or
inversion will significantly reduce this degeneracy since the response
of polarisation CMB spectra is phase-shifted with respect to the total
intensity response.

The total intensity ($T$ modes) of the CMB have been measured
accurately by several instruments
\cite{wmap5,acbar,quad,Jones:2005yb,Pearson:2002tr} to arcminute
scales. Measurements in the gradient ($E$-mode) and curl like
($B$-mode) polarisation components, and their correlation with the
total intensity $TE$, and $TB$, lag behind due to their lower
amplitude. However a number of experiments are now measuring $E$-mode
polarisation with increasing signal-to-noise, starting with the first
detection of $EE$ spectrum \cite{Kovac:2002fg} and subsequent
measurements
\cite{Pearson:2002tr,Montroy:2005yx,wmap5,quad,Bischoff:2008wa}. $B$-mode
polarisation has yet to be measured. Primordial tensor fluctuations
may have an impact on the reconstruction of $P(k)$ on large
scales. However, $B$-modes have not been detected yet and we neglect
their contribution in this work.

We show in this paper that when information contained in both total
intensity and polarisation radiation transfer functions is used in the
reconstruction of the primordial power spectrum tighter constraints
can be obtained. The paper is organised as follows; in section
\ref{rld} we introduce the extension of the Richardson-Lucy algorithm
used to estimate $P(k)$. In Section \ref{fd} we test the algorithm
using simple forecasts of CMB data. Our tests include template input
models with radically broken scale invariance. We explore current
limits on $P(k)$ in section \ref{current} and conclude in section
\ref{diss}.

\section{An extended Richardson-Lucy $P_k$ Estimator}\label{rld}

Direct primordial power spectrum reconstruction requires the inversion
of the following relations
\begin{equation}\label{eq:intro-aps}
C_\ell^{XY} =\int\limits_{-\infty}^{\infty} \frac{\rm{d}k}{k} \Delta_{\ell}^{X}(k) \Delta_{\ell}^{Y}(k) P(k),
\end{equation}
where $X$ and $Y$ represent $T$, $E$, or $B$-type anisotropies,
$C_\ell^{XY}$ are the angular power spectra for the $XY$ combination
and the $\Delta^{X}_{\ell}(k)$ are the photon perturbation transfer
functions. The transfer functions are obtained by integrating the full
Einstein-Boltzmann system of differential equations. These describe the
evolution of perturbations in the photon distribution functions in the
presence of gravity and other sources of stress--energy. The functions
determine all of the structure in the anisotropy spectra which arises
after the initial conditions are set. Most notably the $C_\ell^{XY}$
contain distinct peaks due to the acoustic oscillation of the tightly coupled
photon-baryon fluid in gravitational potential wells at the time of
last scattering. The aim of any inversion method is to distinguish
such features from any structure in the initial perturbation spectrum.

For a finite sampling of the wavenumber space $k$
Eq.~(\ref{eq:intro-aps}) can be recast as an operator acting on the
primordial spectrum $P_k$
\begin{equation}\label{eq:rld-start}
C_\ell = \sum\limits_k F_{\ell k} P_k,
\end{equation}
with operator
\begin{equation}
 F^{X Y}_{\ell k} = \Delta \ln k\,  \Delta_{\ell k}^{X} \Delta_{\ell k}^{Y},
\end{equation}
where $\Delta \ln k$ are the logarithmic $k$ intervals for the chosen
sampling.

A solution for $P_k$ cannot be obtained from a direct inversion of the
$F^{X Y}_{\ell k}$ as it is numerically singular. This is due to the
high level of degeneracy in the transfer functions relating the power
at any wavenumber $k$ to angular multipoles $\ell$ although the system
can be inverted by binning or smoothing appropriately to reduce the
degeneracy \cite{future}.

An alternative approach is the iterative inversion employing the
Richardson-Lucy (RL) method \cite{Richardson, Lucy} for image
reconstruction. The RL method has been widely used in enhancing
telescope images \cite{Jorissen:2001jn,Surpi:2001va,Helder:2008eb} and
it can be shown that the RL method converges to the maximum likelihood
estimator in the case of a Poisson distributed signal
\cite{4307558}. In the following we outline the use of the RL
estimator in reconstructing the primordial power spectrum as
introduced by Shafieloo {\it et al.} \cite{Shafieloo:2003gf} and
extend it to include properly weighted contributions from polarisation
measurements.

Consider the case where the original source plane is the isotropic, primordial
Fourier space spanned by the wavenumber $k$ and the convolved image plane is
the space of angular multipoles $\ell$. In this case the convolution
filter is $F_{\ell k}$ which relates modes the source power $P_k$ to
the image power $C_\ell$.

The RL method provides an iterative solution to (\ref{eq:rld-start})
for $P_k$, given an observed $C_\ell^\textrm{obs}$ with
\begin{equation}\label{eq:rld-basic}
 P_k^{i+1}=P_k^i \sum\limits_{\ell = \ell_{\rm min}}^{\ell_{\rm max}} \tilde{F}_{\ell k} \frac{C_\ell^\textrm{obs}}{C_\ell^i},
\end{equation}
where $C_\ell^i$ is the image obtained from the $i^{\rm th}$ iteration
$P_k^i$ and $\tilde{F}_{\ell k}=F_{\ell k}/\sum_\ell F_{\ell k} $ such that
in the limit $C_\ell^i \rightarrow C_\ell^\textrm{obs}$ we have $
(P_k^{i+1}-P_k^i) \rightarrow 0$. It is also
convenient to recast (\ref{eq:rld-basic}) as change in the $P_k$ \cite{Shafieloo:2003gf}
\begin{equation}\label{eq:rld-cutoff}
P_k^{i+1} =P_k^i\left[ 1+\sum\limits_{\ell=\ell_{\rm min}}^{\ell_{\rm
      max}} \tilde{F}_{\ell
    k}\frac{C_\ell^\textrm{obs}-C_\ell^i}{C_\ell^i}\right],
\end{equation}
such that the cut-offs in multipole space $\ell_{\rm min}$ and
$\ell_{\rm max}$ are not propagated to the iterated solution through
the broadening action of $\tilde F_{\ell k}$.

As it is (\ref{eq:rld-cutoff}) does not account for errors in the
observed image as it applies uniform weighting to all
$C_\ell^\textrm{obs}$. Since the RL estimator is not a well defined
Maximum Likelihood estimator there is not a single choice of optimal
weighting. Observational noise must therefore be included through an
empirically chosen weighting function. We chose to add a weighting
\begin{equation}\label{eq:rld-gk}
G_k=\frac{\sum\limits_\ell \tilde{F}_{\ell k}C_\ell^i}{\sum\limits_\ell \tilde{F}_{\ell k}\left[C_\ell^i+\sigma_\ell \right]},
\end{equation}
where $\sigma_\ell$ is the reported error in the
$C_\ell^\textrm{obs}$. The weights $G_k$ have the properties that if
$\sigma_\ell \ll C_\ell^i$ then $G_k \rightarrow 1$ and if
$\sigma_\ell \gg C_\ell^i$ then $G_k \rightarrow 0$. The weighted
estimator (\ref{eq:rld-cutoff}) now becomes
\begin{equation}\label{eq:rld-error}
  P_k^{i+1} =P_k^i\left[ 1+G_k\sum\limits_\ell \tilde{F}_{\ell k}\frac{C_\ell^\textrm{obs}-C_\ell^i}{C_\ell^i}\right].
\end{equation}

Most CMB experiments observe only part of the sky which leads to
correlated $C_\ell$. The estimator does not account for the
correlations as it includes only diagonal estimates of the uncertainty
in $C_\ell^\textrm{obs}$. Indeed even full-sky experiments such as WMAP
have correlated $C_\ell$ due to galaxy and source cuts. Experiments
observing only small fractions of the sky usually report results in
as a set of bandpowers $C_b = \sum_\ell W^b_\ell C_\ell$, where
$W^b_\ell$ are the bandpower window functions. The bandpowers are less
correlated than individual multipoles and allow us to apply the RL
estimator to cut--sky experiments by calculating band filtered
transfer function operators
\begin{equation}
  C_b = \sum_{\ell k} W^b_\ell\, F_{\ell k} \,P_k \equiv \sum_k F_{bk}\,P_k.
\end{equation}

Using this we can define a generalised iterative estimator with
contributions from any number of bandpowers $b$ as
\begin{equation}\label{eq:banded}
P_k^{i+1}= P_k^i\left[1+ G_k\sum\limits_b \tilde{F}_{b k}\frac{C_b^\textrm{obs}-C_b^i}{C_\ell^i}\right],
\end{equation}
where the $W^b_\ell \rightarrow \delta_{\ell\ell'}$ in the full--sky
limit (WMAP) and are the reported bandpower window functions for other
experiments. In this work we will be using bandpowers from ACBAR
\cite{acbar}, QUaD \cite{quad}, BOOMERanG \cite{Jones:2005yb} and CBI
\cite{Pearson:2002tr} which increase the range of scales probed past
WMAP's resolution limits.

\begin{figure*}[t]
\centering
\includegraphics[width=12cm,angle=270]{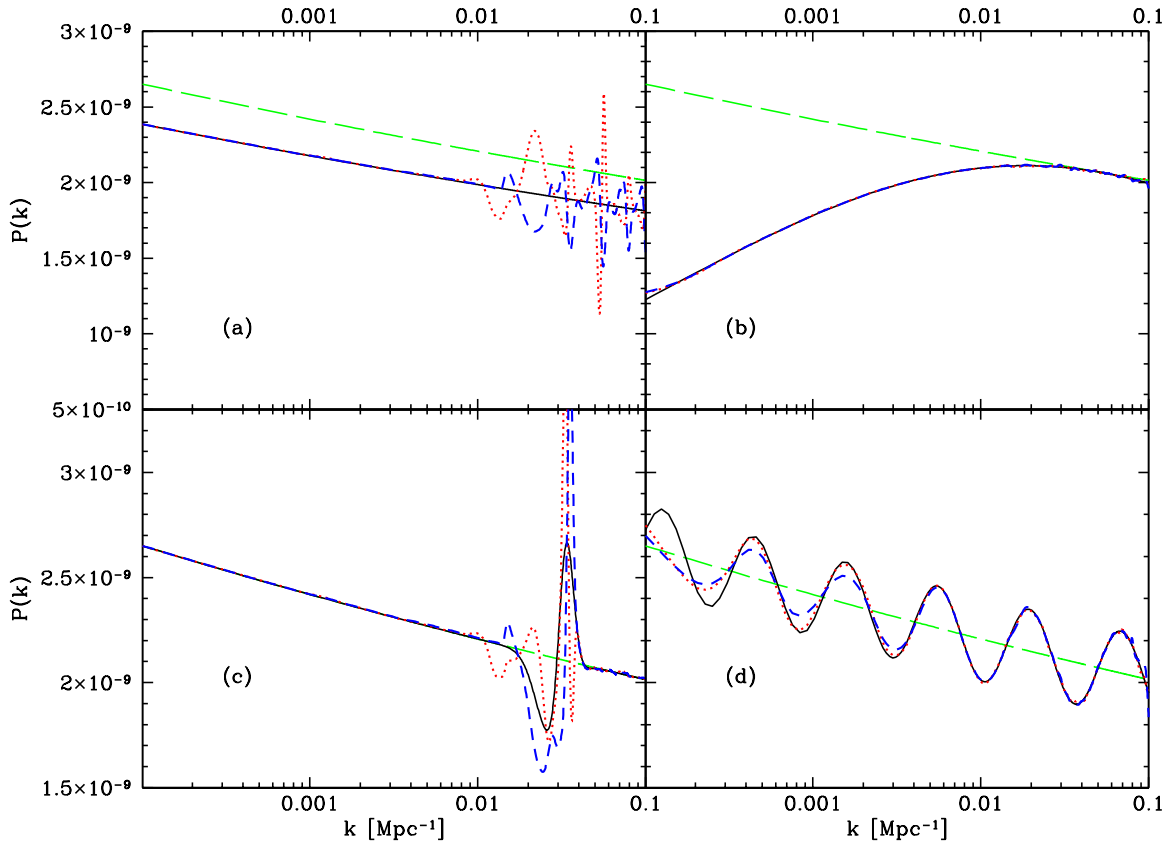}
\caption{The reconstruction of several test spectra. The test models
  used to generate the simulated $C_\ell$, shown in black (solid)
  curves, are (a) A $10$\% decrease in power from the WMAP5
  best fit amplitude, (b) the WMAP5 best fit model including running
  $dn_s/d\ln k = -0.037$, (c) a localised feature at around
  $k=0.02$~Mpc$^{-1}$, and (d) a model with sinusoidal oscillations
  superimposed on the best fit power law spectrum. The green (solid)
  curves show the best fit spectrum used as initial guess in the
  iteration. The red (dotted) curves are the converged reconstructions
  using only total intensity data whereas the blue (dashed) curves use
  both total intensity and polarisation data. The $C_\ell$ forecasts
  assumed an experiment with similar properties to Planck.}
\label{fig:fd-testpk}
\end{figure*}

\begin{figure*}[th]
\centering
\includegraphics[width=12cm,angle=270]{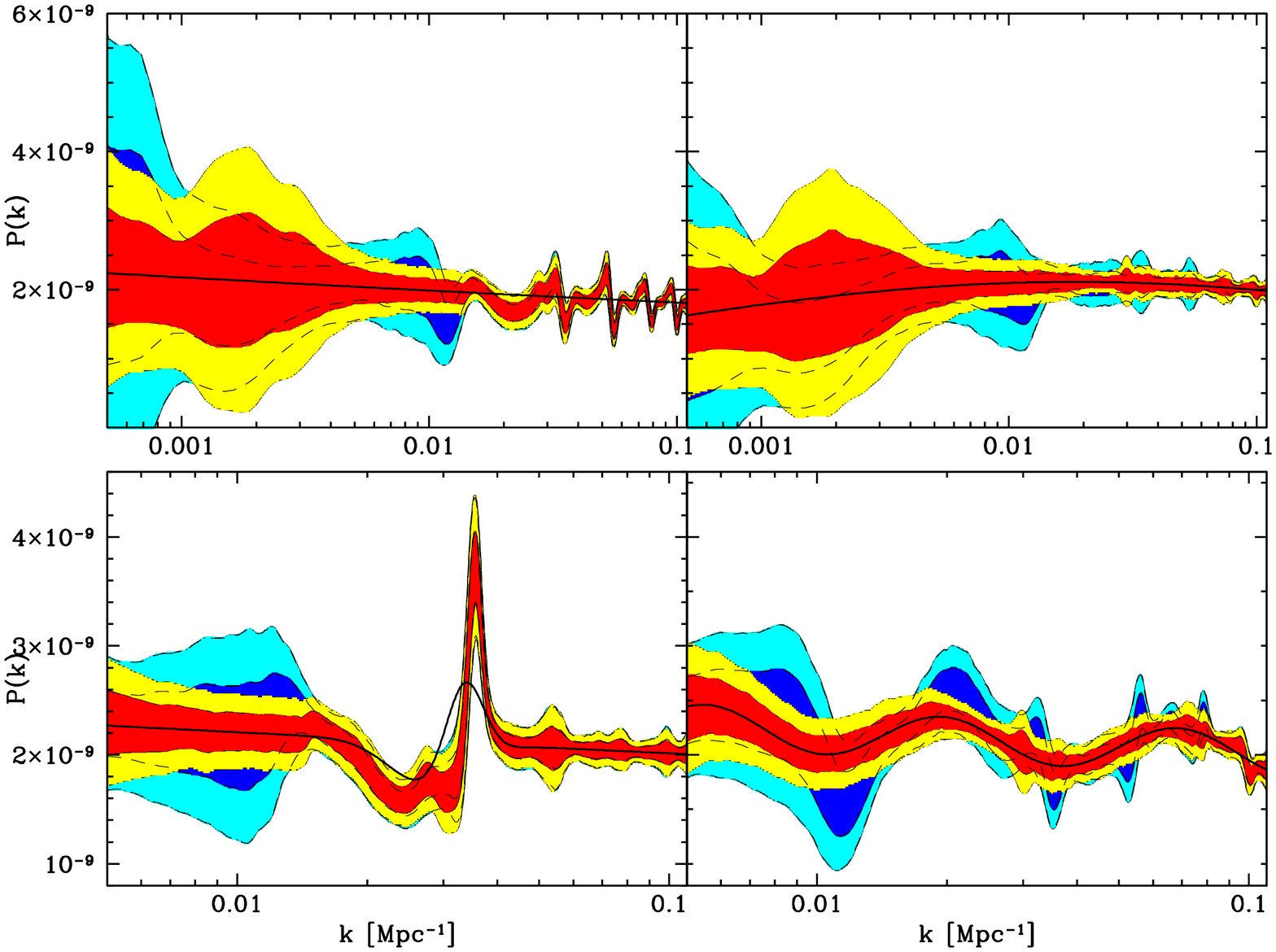}
\caption{Confidence regions around the reconstructed $P_k$ for the
  same test models shown in Fig.~\ref{fig:fd-testpk}. The shaded areas
  indicate the 1 and 2-$\sigma$ confidence regions obtained from 1000
  Monte Carlo realisations of the observed $C_\ell$. The red (solid
  contours) shows the result for polarisation data included overlaid
  on the blue (dashed contours) showing the result for only $TT$
  included. The inclusion of polarisation has the largest impact in
  the case with superimposed oscillations.}
\label{fig:fd-testpk-errors}
\end{figure*}

A problem faced by the RL estimator when including polarisation
measurements is that the polarisation bandpowers are correlated with
the total intensity ones and cannot be included as further linear
contributions to the sum in (\ref{eq:banded}). Following the standard
procedure for extending Maximum Likelihood estimators
to polarised bandpowers we treat each multipole measurement as a
$2\times2$ matrix of bandpowers with   
\begin{equation}
{\bf C}_b = \sum\limits_\ell \left( \begin{array}{cc}
 W_\ell^{b\,TT}C_\ell^{TT} &  W_\ell^{b\,TE}C_\ell^{TE} \\
 W_\ell^{b\,TE}C_\ell^{TE} &  W_\ell^{b\,EE}C_\ell^{EE} \end{array}\right),
\end{equation}
and similarly elevated the transfer function operator to a $2\times2$ matrix
\begin{equation}
{\bf F}_{bk} = \sum\limits_\ell\left( \begin{array}{cc}
 W_\ell^{b\,TT}F_{\ell b}^{TT} &  W_\ell^{b\,TE}F_{\ell b}^{TE} \\
 W_\ell^{b\,TE}F_{\ell b}^{TE} &  W_\ell^{b\,EE}F_{\ell b}^{EE}
 \end{array}\right),
\end{equation}
and we have disregarded any $B$-mode contribution since current
observational limits do not warrant its inclusion. We can now extend
the RL estimator to include the full polarisation information as
\begin{eqnarray}\label{eq:rld-final}
P_k^{i+1} &=&P_k^i{\Bigg (} 1+\nonumber\\
&&\left.{\rm Tr}\left[ {\bf G}_k\cdot \sum\limits_b {\bf
      \tilde{F}}_{b k}\cdot \left({\bf C}_b^\textrm{obs}-{\bf C}_b^i\right)
    \cdot \left({\bf C}_b^i\right)^{-1} \right] \right),\nonumber\\
&&
\end{eqnarray}
where ${\bf G}_k$ is obtained by modifying (\ref{eq:rld-gk}) to sum
over banded quantities and ${\bf \tilde{F}}_{b k}$ is normalised so
that the sum over its diagonal elements is unity.

We have implemented the estimator (\ref{eq:rld-final}) to reconstruct
$P_k$ over a discretely sampled grid in $k$ from a combination of CMB
measurements listed above and for forecasted future data as described
below. The full iteration typically takes ${\cal O}(10)$ minutes on a
current desktop CPU. As a convergence criterion we iterate until the
fractional change in the $P_k$ solution is less than 0.01. Typically
this takes ${\cal O}(100)$ iterations to achieve. We reconstruct $P_k$
over some 3000 $k$ points between $6.9\times 10^{-6}$ Mpc$^{-1}$ and
$0.55$ Mpc$^{-1}$. The sampling is initially logarithmic and switches
to linear at $k=1.25\times 10^{-3}$ Mpc$^{-1}$ following the typical
setup in Einstein-Boltzmann solvers such as {\tt CAMB} \cite{camb} but
with higher resolution.

As shown in \cite{Shafieloo:2003gf} the RL method requires a smoothing
of the recovered $P_k$. We have found that applying a smoothing kernel
at each iteration improves the convergence when using unbinned
$C_\ell$ such as the WMAP results in the estimate. The smoothing helps
to damp down large fluctuations at each iteration driven by the scatter
in the observed data, due to either sample or noise variance. The
behaviour is due to the estimator not being a well defined maximum
likelihood one and thus not taking into account the full correlation
structure of the data in its weighting. We have chosen a simple
smoothing kernel for the applications reported in this work by taking
a nearest neighbour running average over the $P_k$ between each
iteration. Given our $k$ sampling this defines a smoothing size of
$\Delta \log{k}\approx 0.04$ Mpc$^{-1}$ at low $k$ and $\Delta k
\approx 2\times 10^{-4}$ Mpc$^{-1}$ at high $k$. Other smoothing
kernels may lead to more efficient behaviour of the estimator such as
faster convergence but would be more computationally demanding and
would lead to a further loss of resolution in $k$ space.

Since the estimator requires ${\cal O}(10)$ minutes to converge it is
not currently possible to employ it as part of a full Markov
Chain Monte Carlo parameter search in place of the standard primordial
spectrum parameter; amplitude $A_s$ and spectral tilt $n_s$. This may
be possible in future if the algorithm is optimised and/or
parallelised but in this work we chose to explore the $P_k$ ``source''
plane assuming a set of fixed fiducial transfer functions based on
the best-fit parameters. The transfer functions are obtained from {\tt
  CAMB} with parameters $\Omega_bh^2 = 0.0226$, $\Omega_ch^2 = 0.108$,
$\theta = 1.041$, and $\tau = 0.076$.

To estimate the confidence limits around the sampled $P_k$ solution we
Monte Carlo the iterative estimator by generating 1000 simulated sets
of $C_\ell^{\rm obs}$. For simplicity we assume Gaussian distributions for all
the $C_\ell^{\rm obs}$ since more accurate distributions are not
easily reconstructed from some of the published data. We iterate to
convergence on each of the simulated data sets and obtain a covariance
matrix for the samples by averaging over the ensemble
\begin{equation}
  \sigma_{kk'}^2 = \frac{1}{N-1} \sum_{n=1}^{n=N} (P_k^n-\langle P_k\rangle)(P_{k'}^n-\langle P_{k'}\rangle),
\end{equation}  
with $\langle P_k\rangle = \sum_{n=1}^{n=N}P_k/N$. We quote 1 and
2-$\sigma$ errors by taking multiples of the square root of the
diagonal elements of the $\sigma_{kk'}^2$ matrix.

\section{Application of  the extended RL estimator}

\subsection{Tests on forecasted CMB data}\label{fd}

We have tested the extended RL estimator on a number of input $P_k$
templates using forecasted data with similar experimental properties
as the upcoming Planck satellite mission \cite{planck}. We assume a
total of 12 detectors with NETs of 64$\mu K/\sqrt{s}$ observing 80\%
of the sky over 12 months with a resolution of 7 arcminutes FWHM. We
calculate errors around our fiducial CMB best--fit models in both total
intensity and polarisation spectra for this experimental setup and use
these together with $C_\ell$ samples on the fiducial model to test the
estimator's convergence. We consider multipoles of $\ell < 1900$ for
total intensity spectra and $\ell < 1000$ for polarisation. We have
not taken into account any residual error from foreground subtraction
in our forecasts. Thus our forecast are on the optimistic side of the
accuracy achievable, particularly in the polarisation where foreground
removal will certainly have a significant impact on errors at
$\ell < 1000$.

We use four separate $P_k$ templates to generate the $C_\ell$s. The
first is a standard power law with $n_s=0.96$ but with a lower
amplitude than the best--fit, the second is a running $n_s$ model with
$dn_s/d\ln k = -0.037$, the third is a power law with a sharp,
compensated feature at $k=0.02$~Mpc$^{-1}$ \cite{Lesgourgues:1999uc,Adams:2001vc,Feng:2003zua,Mathews:2004vu,Joy:2007na,Hunt:2007dn,Jain:2008dw,Lerner:2008ad} and the fourth is a power
law with superimposed sinusoidal oscillations \cite{Wang:2002hf,Martin:2003sg,Martin:2004yi,Pahud:2008ae,Romano:2008rr}. The extended estimator
is run on all four sets of $C_\ell$ forecasts and the resulting $P_k$
solution is compared to the input one. In Fig.~\ref{fig:fd-testpk}
we show the results for runs including $C_\ell^{TT}$ only and runs
including $C_\ell^{TT}$, $C_\ell^{TE}$, and $C_\ell^{EE}$. 

Each run is started with a first guess $P_k$ (green/long--dashed)
which is the current best--fit power law spectrum. In all cases the
structure in the input $P_k$ is reproduced to some degree over a large
range in $k$. Beyond the $k$ range shown in Fig.~\ref{fig:fd-testpk} there are
significant departures from the correct solution. On the large scales
($k <0.0001$~Mpc$^{-1}$) this is due to the cut--off in the transfer
functions due to the horizon scale. At small scales
($k>0.1$~Mpc$^{-1}$) the cut--off is due to the resolution limit of the
forecasts. We find the method is particularly well suited for
reconstructing long wavelength structure such as in the running and
oscillating model. Although the sharp feature in the third model is
present in the reconstructed spectrum, its amplitude is not recovered
accurately. This example highlights the limitations of such methods in
reconstructing high frequency features in the primordial spectrum. In
the first model we also see high frequency features on small scales,
this is due to the difference in amplitude between the initial guess
and correct solution. A power law initial guess with closer starting
amplitude to the correct value reduces this small scale noise. This
could be easily obtained by carrying out a standard power law fit to
the data before running the reconstruction estimate.

In Fig.~\ref{fig:fd-testpk-errors} we show the same set of
reconstructions but include 1 and 2-$\sigma$ confidence regions
obtained from the Monte Carlo covariance matrix. We have over sampled
the $P_k$ and care should be taken in interpreting the significance of
any feature as the samples are highly correlated. The plot includes a
set of contours for the reconstruction using only total intensity data
(blue/dashed) and including polarisation (red/solid). The range in
$k$ has been modified to emphasise the regions where the best limits
are obtained. The addition of polarisation data gives additional
constraints on the smaller scales particularly in the oscillating
model case. This is not surprising since total intensity and
polarisation measurements are complementary in differentiating any
structure in the primordial spectrum from acoustic oscillations
imprinted on the CMB at last scattering. The polarisation data has
little effect in improving the constraints on the spectrum with a
compensated feature although a detection of the feature is evident.

\subsection{Current Limits}\label{current}

In this section we use the extended RL estimator to reconstruct the
primordial power spectrum using currently available data. Our
combination of data includes the latest WMAP release \cite{wmap5}
including $TT$ for $\ell < 700$, the CBIpol results
\cite{Pearson:2002tr} including $TT$ band powers 2 to 12 and all
polarisation band powers, the BOOMERanG 2003 \cite{Jones:2005yb}
flight band powers, the QUaD \cite{quad} results, and the ACBAR 2008
\cite{acbar} excluding the last two band powers due to the excess
power. We have excluded the WMAP polarisation measurements as they are
too noisy to be used in the reconstruction would require binning to
reduce the scatter. 

\begin{figure}
\centering
\includegraphics[width=9.0cm]{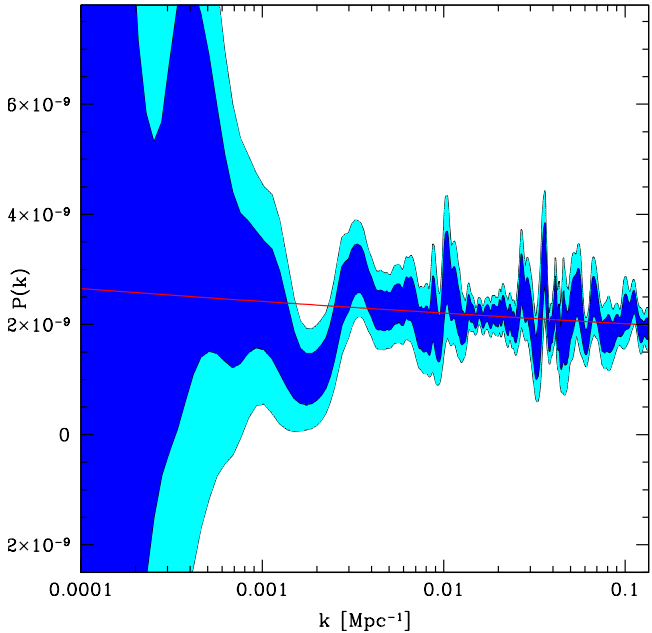}
\caption{Current limits from a combination of CMB data sets (WMAP,
  ACBAR, QUaD, BOOMERanG and CBI). There is some evidence of a dip in
  power at around $k\approx0.002$ below the best fit power law
  model. Shaded regions are defined as in Fig.~\ref{fig:fd-testpk-errors}.}
\label{fig:current-all}
\end{figure}

The overall effect of adding in the sub--orbital experiment is to
extend the $k$ range of the reconstruction past $k\sim
0.07$~Mpc$^{-1}$ where the WMAP noise becomes too large for the data
to be included in the estimator. Given WMAP's resolution in multipole
space, it strongly dominates the reconstruction on scales larger than
$k\sim 0.07$~Mpc$^{-1}$. In Fig.~\ref{fig:current-all} we show the
results of the reconstruction. There is no indication of a departure
from a pure power law for the recovered spectrum except for a dip at
$k\sim 0.002$~Mpc$^{-1}$.

\begin{figure}[t]
\centering
\includegraphics[width=9.0cm]{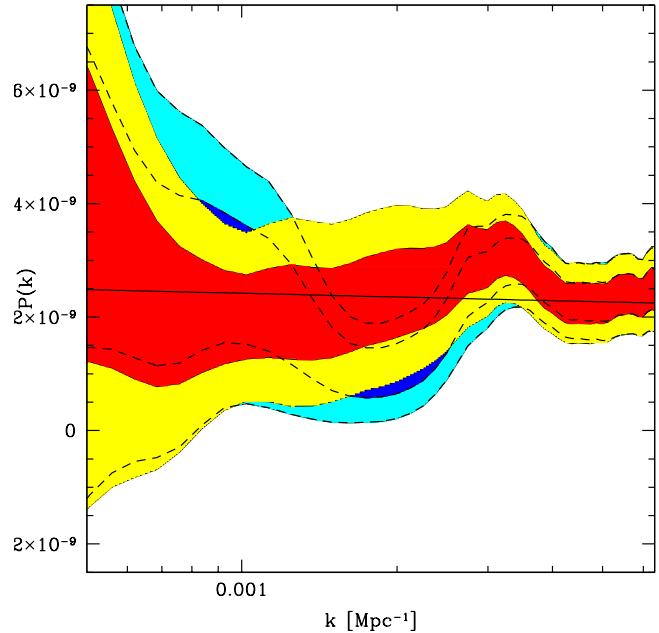}
\caption{As an indication of the origin of the dip at $k\approx0.002$ Mpc$^{-1}$ we
  remove the data between $\ell=18$ and $\ell=26$ and re-run the
  estimator. The red/yellow contours show the effect of the removal
  over the original estimate (blue/cyan).}
\label{fig:current-cut1826}
\end{figure}

To determine the origin of the feature at $k\sim 0.002$~Mpc$^{-1}$ we
re-estimate the power spectrum with a cutout of the WMAP data in the
range $18 \le \ell \le 26$ where the $C_\ell$ lie significantly below
the best--fit models. Fig.~\ref{fig:current-cut1826} shows the result
of for the reconstruction. No evidence for the feature remains when
the $18 \le \ell \le 26$ WMAP data is removed from the estimate
indicating that the dip is associated with the feature in the
$C_\ell$. 

\begin{figure}[t]
\centering
\includegraphics[width=9.0cm]{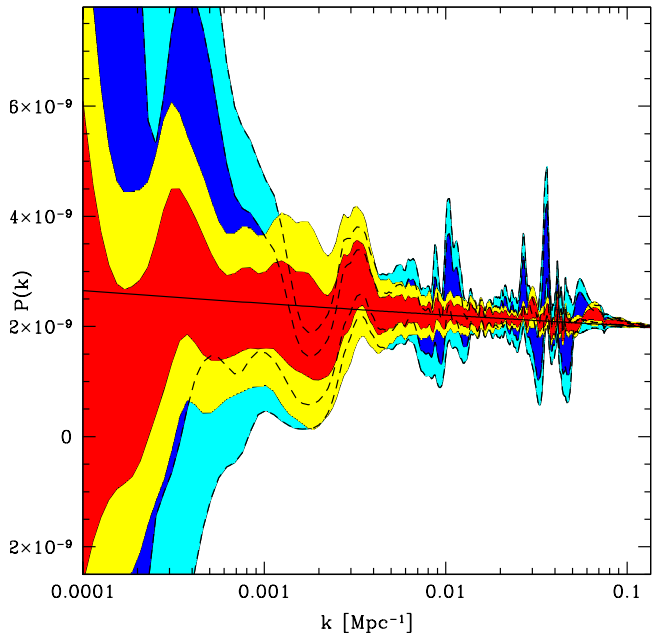}
\caption{We show the improvement that polarisation data can have on
  the estimation of the spectrum. The blue area is the uncertainty on
  $P_k$ given only the WMAP 5-year data. In the foreground we have
  combined this data with polarisation data from our Planck-like
  experiment and show that it significantly reduces the inherent
  error.}
\label{fig:current-pol}
\end{figure}

\section{Discussion}\label{diss}

The RL estimator provides a model independent method for
reconstruction of the primordial power spectrum of perturbations $P_k$
from measurements of the CMB angular power spectrum $C_\ell$. We have
extended the RL estimator to use on multiple data sets including
properly weighted polarisation data and have used Monte Carlo
realisations of the input $C_\ell$ measurements to estimate confidence
limits around the reconstructed spectra.

We have applied the new estimator to current measurements.These
include band powers from the QUaD experiment which provide the highest
signal--to--noise measurements of $EE$ power so far. Including the
polarisation information increases the constraints on the power
spectrum reconstruction as it carries independent information which is
complementary in phase to that of the total intensity. However current
$TE$ and $EE$ measurements are still noise dominated and do not
contribute significantly but future sample variance limited
measurements will help to constrain any structure in the primordial
spectrum as shown in our examples using forecasted data. An exception
to this is the current WMAP polarisation data which may have some
impact if it were binned to reduce noise scatter. We have not explored
this option in this work as this would have required binning of the
$TT$ data too with a resulting loss of resolution in $k$ space.

We have found that the addition of polarisation data is particularly
helpful in constraining oscillatory structure in $P_k$. Spectra with
superimposed sinusoidal features have been considered in the
literature and have been constrained using model dependent fits
\cite{Martin:2004yi,Hamann:2007pa,Pahud:2008ae,Liu:2009nv}.

Due to the empirical nature of the weighting used the estimator has a
limited acceptance range in the signal--to--noise of the band powers
it is run on. Low signal--to--noise $C^{\rm obs}_\ell$ or $C_b^{\rm
  obs}$ cause the estimator to converge very slowly. Conversely if the
data is weighted too strongly there is a loss of resolution and
signal--to--noise in the reconstructed $P_k$. The best approach is to
exclude the low signal--to--noise measurements from the data being
used.  It is possible
that other weighting schemes may be more efficient in allowing the
whole range of $C_\ell$ measurements to be used. However, in its
present form, the estimator will be most useful when all polarisations
will be sample variance limited out to the same $\ell$. As an example
of the improvement this will have over current estimates we add our
forecasted $TE$ and $EE$ data to the current measurements and run the
extended RL estimator. The result is shown in
Fig.~\ref{fig:current-pol} which shows an overall improvement in the
Monte Carlo confidence limits over the entire range in $k$ being
reconstructed.

In this work we have not explored the degeneracy of the method with
the physical parameters determining the structure of the CMB transfer
functions. The current run--time to convergence for each
reconstruction is too long to allow the tens of thousands of runs
required to implement the method as part of a full Markov Chain Monte
Carlo exploration of the entire parameter space. However with
parallelisation and more efficient convergence it may become possible
to do this in the near future.

There are further extensions of the RL estimator that could increase
its effectiveness. The addition of other observables with different
transfer functions such as galaxy redshift surveys or cosmic shear
surveys will provide complementary information in the
reconstruction. Our extension of the RL estimator is compatible with
these other observables as they can be included as further independent
data points with properly formatted and binned transfer
functions. Such extensions would contribute key information in the
reconstruction on scales where the presence of foregrounds in the CMB
reduces its effectiveness.

\acknowledgements

  This work was supported by a STFC studentship and used the Imperial
  College high performance computing
  service\footnote{http://www.imperial.ac.uk/ict/services/teachingandresearchservices/highperformancecomputing}.

\bibliography{paper}

\end{document}